\documentclass[aps,prd,twocolumn,showpacs,preprintnumbers,amsmath,amssymb,groupedaddress]{revtex4}
\usepackage{graphicx}
\usepackage{txfonts}

\begin{document}


\title{Gravitational Lensing Analyzed by Graded Refractive Index of Vacuum}


\author{Xing-Hao Ye}
\email[Electronic address: ]{yxhow@163.com}
\author{Qiang Lin}
\email[Corresponding author: ]{qlin@zju.edu.cn}

\affiliation{Department of Physics, Zhejiang University, Hangzhou
310027, China }


\date{\today}

\begin{abstract}
We found strong similarities between the gravitational lensing and
the conventional optical lensing. The similarities imply a graded
refractive index description of the light deflection in
gravitational field. We got a general approach to this refractive
index in a static spherically symmetric gravitational field and
obtained its exterior and interior solutions exactly through the
general relativity. In weak field case, the two solutions come to a
simple unified exponential function of the gravitational potential.
With these results, the gravitational lensing can be analyzed in a
convenient optical way. Especially, the long puzzling problem of the
central image missing can be solved easily. We also pointed out that
the graded refraction property of the gravitational spacetime is
related to the vacuum influenced by the gravitational matter.
\end{abstract}


\maketitle


\section{Introduction}
\label{}

Gravitational lensing is an effect predicted by general relativity
\cite{Mollerach2002,Wambsganss1998,Ohanian1994}. It has been now a
powerful tool for the study of the so intractable problems in
astrophysics and cosmology such as the value of Hubble constant
\cite{Kundic1997}, the physics of quasars \cite{Solomon2003}, the
mass and mass distribution of a galaxy or a galaxy
cluster\cite{Wambsganss1998}, the large scale structure of the
universe \cite{Massey2007,Springel2006}, the existence of dark
matter \cite{Freeman2003,Inada2003,Wittman2000}, the nature of dark
energy \cite{Bennett2006,Bennett2005} and so on.

It is told that, the light deflection of a gravitational lens is
caused by the gravity or in Einstein's words the curved spacetime,
while the light deflection of an optical lens is caused by the
variance of refractive index of the medium. But is there any deeper
similarity between the two lenses besides the deflection of light?
Can we analyze the gravitational lensing as the conventional optical
one?

In fact, people have long tried an optical description of the
lensing effect. In 1920, Eddington \cite{Eddington1920} suggested
that the light deflection in solar gravitational field can be
conceived as a refraction effect of the space in a flat spacetime.
The idea was further studied by Wilson \cite{Wilson1921}, Dicke
\cite{Dicke1957}, Felice \cite{Felice1971}, and Nandi \emph{et al}.
\cite{Nandi1995,Evans1996,EvansNandi1996}. Recently, this thought of
light deflection has been investigated further by Puthoff
\cite{Puthoff2002,Puthoff2005}, Vlokh \cite{Vlokh2004-2006}, Ye and
Lin \cite{Ye2007} etc. In Puthoff's paper, the light deflection is
related to the vacuum polarization in gravitational field. Vlokh
discussed further this light refraction effect. Ye and Lin used the
refractive index to simulate the gravitational lensing.

The purpose of this paper is, on the basis of the general
relativity, to find the similarities between the gravitational
lensing and the optical lensing, and to analyze the former in a
simple optical way. The paper is organized as follows: in Sec. II,
we point out the strong similarities between the two lenses and
suggest a refractive index analysis of the gravitational lensing; in
Sec. III, we find the exact solutions of this refractive index for
outside and inside the lens matter system respectively; in Sec. IV,
we make a weak field approximation; in Sec. V, we apply the obtained
result to the problem of gravitational lensing, especially to the
central image missing problem puzzled physicists for a long time
\cite{Winn2004}; in Sec. VI, we make a discussion on the refraction
property of the gravitational spacetime; finally in Sec. VII, we
draw our conclusions.

\section{Similarities between the two lenses} \label{}

\subsection{Similarity in Fermat's principle } \label{}

Landau and Lifshitz have derived from the general relativity the
Fermat's principle for the propagation of light in a static
gravitational field as follows \cite{Landau1975}:
\begin{equation}
\delta \int {g_{00}}^{-1/2}dl=0,
\end{equation}
where $dl$ is the length element of the passing light measured by
the local observer, $g_{00}$ is a component of the metric tensor
$g_{\mu\nu}$, $g_{00}^{-1/2}dl$ corresponds to an element of optical
path length. ${g_{00}}^{-1/2}=dt/d\tau$, where $d\tau$ represents
the time interval measured by the local observer for a light ray
passing through the length $dl$, while $dt$ is the corresponding
time measured by the observer at infinity. Eq.\ (1) could then be
rewritten as
\begin{equation}
\delta \int \frac{dt}{d\tau}\frac{dl}{ds} ds = 0,
\end{equation}
where $ds$ is the length element measured by the observer at
infinity, corresponding to the local length $dl$.

For an optical lens, the light propagation satisfies the
conventional Fermat's principle
\begin{equation}
\delta \int n ds=0,
\end{equation}
where $n$ is the refractive index of the medium.

The similarity between Eq.\ (2) and Eq.\ (3) indicates that if we
set the scale of length and time at infinity as a standard scale for
the whole gravitational space and time, the light propagating in a
gravitational lens could then be regarded as that in an optical
medium with the refractive index being
\begin{equation}
n=\frac{dt}{d\tau} \frac{dl}{ds},
\end{equation}
where $dt/d\tau$ relates to the curved time and $dl/ds$ relates to
the curved space.

\subsection{Similarity in light deflection formula} \label{}

First we consider the light deflection in a gravitational lens. For
a static and spherically symmetric lens matter system, the metric
has the standard form
\begin{equation}
d\mathcal {T}^2=B(R) c^2 dt^2-A(R)dR^2-R^2(d\theta^2+sin^2\theta
d\phi^2),
\end{equation}
where $R$ is the radial coordinate of the metric.

The light deflection is shown in Fig.\ 1, where curve
$\textnormal{P}_0\textnormal{P}$ represents the light ray, $M$ is
the mass of the lens matter, $\theta$ is the angular displacement of
the coordinate radius $R$, $\beta$ is the angle between the
coordinate radius $R$ and the tangent of the ray at point P. The
general relativity gives the angular displacement as follows
\cite{Arceo2006,Virbhadra2002,Weinberg1972}:
\begin{equation}
d\theta=\frac{dR}{R/\sqrt{A(R)}
\sqrt{\left[\frac{R/\sqrt{B(R)}}{R_0/\sqrt{B(R_0)}}\right]^2-1}},
\end{equation}
where $R_0$ represents the radial coordinate at the nearest point
$\textnormal{P}_0$ .

Now we consider the light deflection in a medium of a spherically
symmetric refractive index $n$. According to the Fermat's principle,
the light propagation in such a lens satisfies the following
relation \cite{Wolf1999}:
\begin{equation}
n r \sin{\beta}=\textnormal{constant},
\end{equation}
where $r$, differing from the above said coordinate radius $R$, is
the distance from the light to the center of the lens. The relation
can be rewritten as
\begin{equation}
n r \sin{\beta}=n_0 r_0,
\end{equation}
where $r_0$ and $n_0$ represent the radial distance and the
refractive index at the point closest to the center respectively.

\begin{figure}
\includegraphics[width=2.0in]{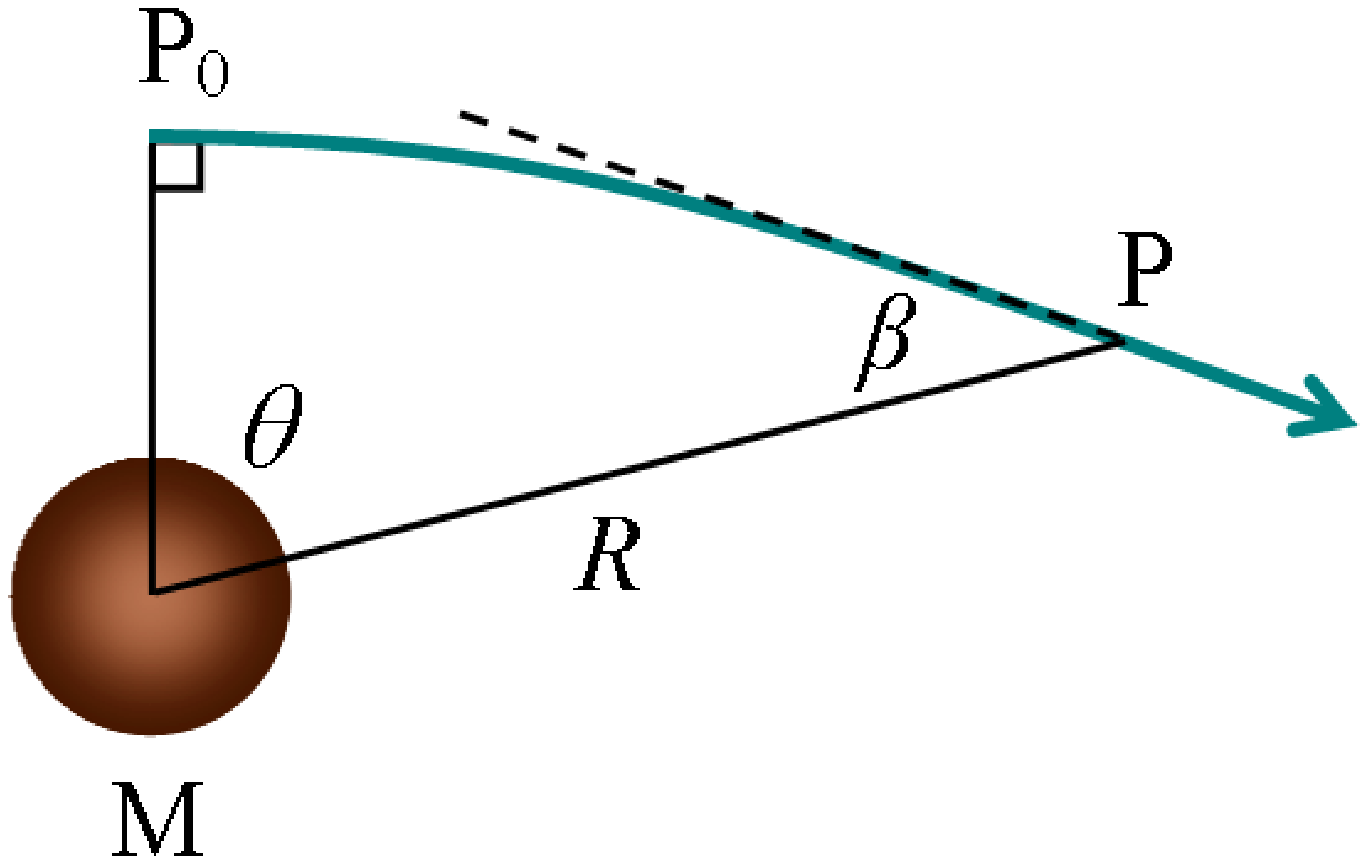}
\centering \caption{\label{fig01} Light deflection in a
gravitational lens.}
\end{figure}

Since
\begin{equation}
\tan{\beta}=\frac{rd\theta}{dr},
\end{equation}
associating with Eq.\ (8) reaches
\begin{equation}
d\theta=\frac{dr}{r \sqrt{\left(\frac{nr}{n_0 r_0}\right)^2-1}}.
\end{equation}

Eq.\ (10) and Eq.\ (6) show another strong similarity between the
two lenses, which once more indicates that there could be a
refractive index analysis of the gravitational lensing. The
similarity tells us that this refractive index can be figured out
through the following two equations:
\begin{eqnarray}
\frac{dR}{R/\sqrt{A(R)}}=\frac{dr}{r},\\
n=\frac{R}{r \sqrt{B(R)}}.\ \ \
\end{eqnarray}

For the detailed derivation of the above two equations, see the
APPENDIX.

\section{Exact solutions of the refractive index}
\label{}

\subsection{Exterior solution}
\label{}

 The coefficients $A(R)$ and $B(R)$ in Eqs.\ (11) and (12) can be obtained from the
Schwarzschild solutions. The Schwarzschild exterior solution ($R
\geqslant R_L$, $R_L$ is the radial coordinate at the surface of the
lens matter system) gives \cite{Weinberg1972}:
\begin{eqnarray}
A(R)=\left(1-\frac{2GM}{Rc^2}\right)^{-1},\\
B(R)=1-\frac{2GM}{Rc^2},\ \ \ \ \
\end{eqnarray}
where $G$ is the gravitational constant, $c$ is the velocity of
light in vacuum without the influence of gravitational field.

Substituting Eq.\ (13) into Eq.\ (11) gives
\begin{equation}
\frac{r}{k}=\frac{\sqrt{1-\frac{2GM}{Rc^2}}+1}{\frac{2GM}{Rc^2}}-\frac{1}{2},
\end{equation}
where $k$ is a constant. Since in the Schwarzschild metric, the
ratio $r/R \rightarrow 1$ at infinity, we have
\begin{equation}
k=\frac{GM}{c^2}.
\end{equation}
Thus Eq.\ (15) gives
\begin{equation}
r=\frac{R}{2}\left(\sqrt{1-\frac{2GM}{Rc^2}}+1-\frac{GM}{Rc^2}\right),
\end{equation}
or
\begin{equation}
R=r\left(1+\frac{GM}{2rc^2}\right)^2.
\end{equation}
Then through Eqs.\ (12), (14) and (17), we get the exterior
refractive index
\begin{equation}
n=\frac{1}{\frac{1}{2}
\left(\sqrt{1-\frac{2GM}{Rc^2}}+1-\frac{GM}{Rc^2}\right)\sqrt{1-\frac{2GM}{Rc^2}}},
\end{equation}
or through Eqs.\ (12), (14) and (18), we get
\begin{equation}
n=\left(1+\frac{GM}{2rc^2}\right)^3
\left(1-\frac{GM}{2rc^2}\right)^{-1},
\end{equation}
which is exactly in agreement with that given by Felice in a
different way \cite{Felice1971}.

\subsection{Interior solution}
\label{}

The Schwarzschild interior solution ($R \leqslant R_L$) gives
\cite{Weinberg1972}
\begin{eqnarray}
A(R)=\left(1-\frac{2GM(R)}{Rc^2}\right)^{-1}, \ \ \ \ \ \ \ \ \ \ \
\ \ \ \ \
\ \ \ \ \\
B(R)=\textnormal{exp}\ \ \ \ \ \ \ \ \ \ \ \ \ \ \ \ \ \  \ \ \ \ \ \ \ \ \ \ \ \ \ \ \ \ \ \ \ \ \ \ \ \ \ \ \nonumber\\
\left\{ -\int_R^{\infty} {\frac{2G}{R^2 c^2}\left[M(R)+\frac{4\pi
R^3p(R)}{c^2}\right] \left[1-\frac{2GM(R)}{Rc^2}\right]^{-1}dR }
\right\},
\end{eqnarray}
where $M(R)=\int_0^R{4 \pi R^2 \rho(R)dR}$, $\rho(R)$ is the mass
density, and for an ordinary lens matter system, the pressure
$p(R)=0$. So
\begin{eqnarray}
B(R)=\textnormal{exp} \left\{ -\int_R^{R_L} {\frac{2GM(R)}{R^2 c^2}
\left[1-\frac{2GM(R)}{Rc^2}\right]^{-1}dR } \right\}\nonumber
\\
\textnormal{exp}\left\{ -\int_{R_L}^{\infty} {\frac{2GM(R_L)}{R^2
c^2} \left[1-\frac{2GM(R_L)}{Rc^2}\right]^{-1}dR } \right\}.
\end{eqnarray}

Substituting Eq.\ (21) into Eq.\ (11), and considering the relation
between the radial coordinates $r_L$ and $R_L$ at the surface
determined by Eq.\ (17), we obtain
\begin{eqnarray}
r= \frac{R_L}{2}\left(\sqrt{1-\frac{2GM(R_L)}{R_L
c^2}}+1-\frac{GM(R_L)}{R_L
c^2}\right)\nonumber \\
\textnormal{exp}\left(-\int_{R}^{R_L}\frac{dR}{R
\sqrt{1-\frac{2GM(R)}{Rc^2}}}\right).
\end{eqnarray}

Then combining Eqs.\ (12), (23) and (24) gives the interior
refractive index
\begin{eqnarray}
n= n_L \frac{R}{R_L}\textnormal{exp}\left(\int_{R}^{R_L}\frac{dR}{R
\sqrt{1-\frac{2GM(R)}{Rc^2}}}\right)\nonumber \\
\textnormal{exp}\left(\int_{R}^{R_L}\frac{GM(R)}{R^2c^2}
\left[1-\frac{2GM(R)}{Rc^2}\right]^{-1}dR\right),
\end{eqnarray}
where
\begin{equation}
n_L=\frac{1}{\frac{1}{2} \left(\sqrt{1-\frac{2GM(R_L)}{R_L
c^2}}+1-\frac{GM(R_L)}{R_L c^2}\right)\sqrt{1-\frac{2GM(R_L)}{R_L
c^2}}},
\end{equation}
is the refractive index at the surface.

\section{Weak field approximation}
\label{}

For an ordinary lens matter system, the gravitational field is not
extremely strong, i.e., $GM/Rc^2$ or $GM/rc^2<<1$, then we have:

The exterior refractive index
\begin{equation}
n \cong
\textnormal{exp}\left(\frac{2GM}{rc^2}\right)=\textnormal{exp}\left(\frac{-2P_r}{c^2}\right),\
\end{equation}
where $P_r=-GM/r$ is the gravitational potential at position $r$
outside the lens matter system.

The above expression of graded refractive index has been verified in
the problem of light deflection in the solar gravitational field
\cite{Ye2007}, with the result being in agreement with that given by
the general relativity \cite{Weinberg1972,Ohanian1994} and the
actual measurements \cite{Fomaleont1976}.

And the interior refractive index

\begin{equation}
n \cong \textnormal {exp} \left[\frac{2GM(r_L)}{r_L
c^2}+\int_r^{r_L}{\frac{2GM(r)}{r^2
c^2}dr}\right]=\textnormal{exp}\left(\frac{-2P_r}{c^2}\right),
\end{equation}
where $P_r=-GM(r_L)/r_L-\int_r^{r_L}{[GM(r)/r^2]dr}$ is the
gravitational potential at position $r$ inside the lens matter
system.

In general, we have the refractive index profile both outside and
inside the lens matter system in weak field case as follows:
\begin{equation}
n = \textnormal{exp}\left(\frac{-2P_r}{c^2}\right).
\end{equation}

The above result is derived from a single static spherically
symmetric lens matter system. For a multi-body system, the total
gravitational potential will be the superposition of each potential;
therefore, the refractive index can be expressed as
\begin{eqnarray}
n &=& \textnormal{exp}\left(\frac{-2 P_r}{c^2}\right)\nonumber\\
&=&
\textnormal{exp}\left[\frac{-2(P_{r1}+P_{r2}+P_{r3}+\cdots)}{c^2}\right]\nonumber\\
&=& n_1 n_2 n_3 \cdots,
\end{eqnarray}
where $P_{r1}, P_{r2}, P_{r3}, \cdots$ and $n_1, n_2, n_3, \cdots$
are the gravitational potential and the corresponding refractive
index caused by each gravitational body respectively. This
expression may be extended to arbitrary distributed matter systems.
Fig.\ 2 shows such an example, where the brighter cyanine represents
the higher value of refractive index, and the closed curves are the
isolines --- the denser the lines, the quicker the change of
refractive index.

\begin{figure}
\centering
\includegraphics[width=2.8in]{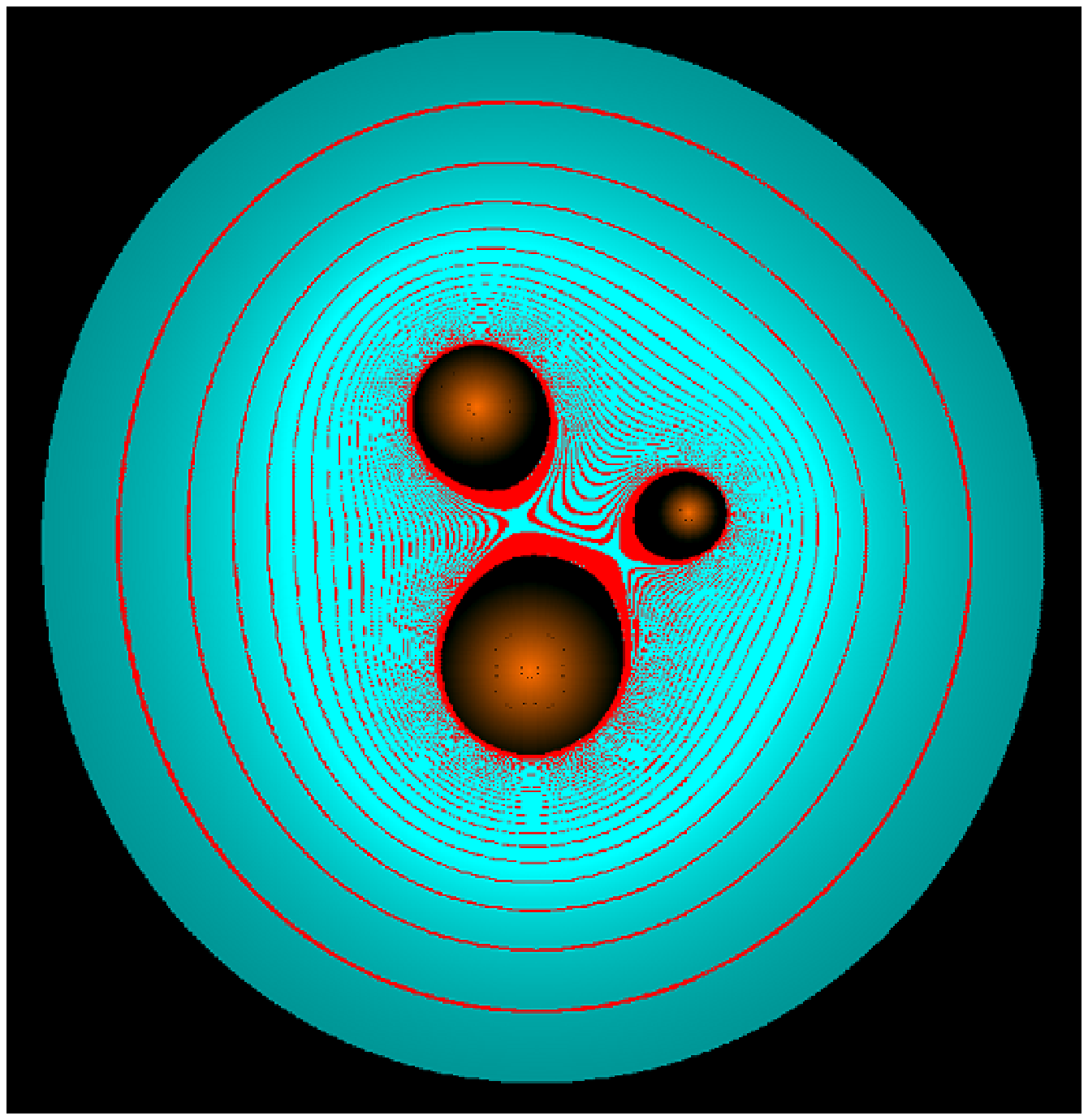}
\caption{\label{fig02} Refractive index profile of a gravitational
lens composed of three celestial bodies of different mass.}
\end{figure}

\section{Application to gravitational lensing}
\label{}

Eq.\ (29) and Eq.\ (30) are the main results of this paper, which
provide a convenient optical way to describe the effect of
gravitational lensing. Considering a source $S$ and a lens $L$ of
mass $M$, the light emitted from $S$ is bent due to the
gravitational field of the lens. The bent light could be figured out
through Eq.\ (7) and Eq.\ (29). Drawing the extension line of the
light from the observer $O$, the apparent (observed) position of the
source image could then be found out. The result is shown in Fig.\
3, where $I_1$, $I_2$ represent the upper and lower images
respectively.

\begin{figure}
\centering
\includegraphics[width=2.8in]{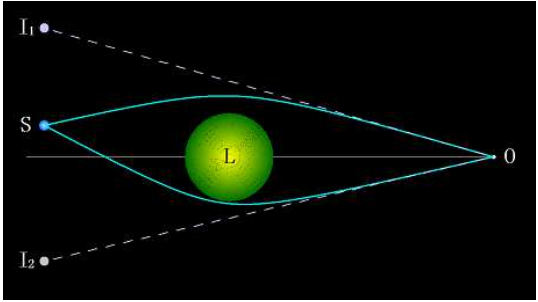}
\caption{\label{fig03}  A computer simulation of gravitational
lensing.}
\end{figure}

The refractive index given by Eq.\ (29) could also be applied in
studying the formation of the central image, which is predicted by
the general relativity but not observed in almost all known cases of
gravitational lensing. This problem has puzzled people for many
years \cite{Winn2004}.

For the central imaging, the refractive index profile inside the
lens matter system should also be considered. As a model for
discussion, we suppose a matter system (for example, a galaxy or a
cluster of galaxies) of radius $r_L$ with a density distribution
given by
\begin{equation}
\rho=\rho_c \left[1-\left(\frac{r}{r_L}\right)^k \right],
\end{equation}
where $\rho_c$ is the central density of the system, $0\leqslant
r\leqslant r_L$, $k>0$. The density $\rho$ decreases with the
distance $r$ from the center of mass; the decreasing rate depends on
the parameter $k$. This model gives the distribution of
gravitational potential as

\begin{eqnarray}
P_{ro}=-4\pi \rho_c
G\frac{k}{3(3+k)}\frac{r_L^3}{r},\ \ \ \ \ \ \ \ \ \ \ \  \ \ \ \ \ \ \ \ \ \ \ \ \ \ \  \\
P_{ri}=-4\pi \rho_c G \left
\{\frac{k}{2(2+k)}r_L^2-\left[\frac{1}{6}-\frac{1}{(2+k)(3+k)}\left(\frac{r}{r_L}\right)^k\right]r^2\right\}
,
\end{eqnarray}
for outside ($r\geqslant r_L$) and inside ($r\leqslant r_L$) the
matter system respectively.

The refractive index profile outside and inside the lens matter
system then reads

\begin{eqnarray}
n_o=\textnormal{exp}\left[\frac{8 \pi \rho_c
G}{c^2}\frac{k}{3(3+k)}\frac{r_L^3}{r} \right],\ \ \ \ \ \ \ \ \ \ \ \ \ \ \ \ \ \ \\
n_i=\textnormal{exp} \ \ \ \ \ \ \ \ \ \ \ \ \ \ \ \ \ \ \ \ \ \ \ \ \ \ \ \ \ \ \ \ \ \ \ \ \ \ \ \ \ \ \ \ \ \ \ \ \ \ \ \ \ \ \nonumber\\
\left\{\frac{8\pi \rho_c G}{c^2}\left \{\frac{k}{2(2+k)}r_L^2-\left
[\frac{1}{6}-\frac{1}{(2+k)(3+k)}\left(\frac{r}{r_L}\right)^k\right]r^2\right\}\right\}.
\end{eqnarray}

\begin{figure}
\centering
\includegraphics[width=2.8in]{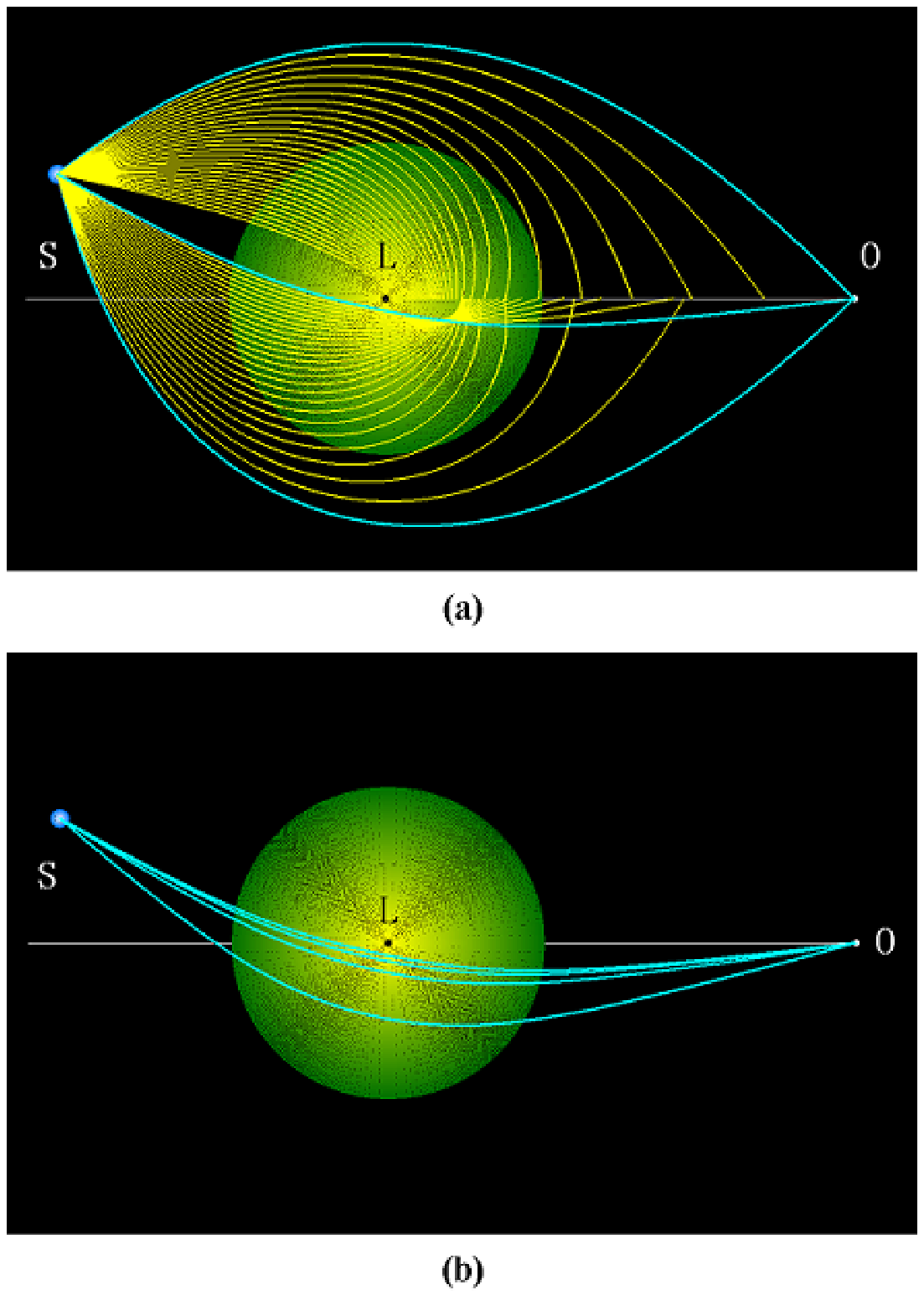}
\caption{\label{fig04}  Ray tracing results for the central imaging.
(a) Tracing the central imaging ray for a given lensing system; (b)
Tracing the central imaging rays for lenses of different mass.}
\end{figure}

Fig.\ 4 (a) shows a ray tracing result for the imaging of a
gravitational lens with the above described refractive index
profile. In the figure, only three paths of ray (indicated by the
three thick lines) could pass through the observer $O$, forming the
upper, lower and central images respectively. From the figure, we
find that, the larger the distance $OL$ from the observer to the
lens body, the closer the central imaging ray to the lens center. In
the same way, the larger the distance $SL$ from the source to the
lens body, the closer the central imaging ray to the lens center.
This can be easily seen by interchanging the source and observer in
the figure.

Fig.\ 4 (b) shows the dependence of the ray position of the central
image on the lens mass. The lens mass is expressed by
$M=\frac{4}{3}\pi r_L^3 \rho_c k/(3+k)$, which can be derived from
Eq.\ (31). The four curved lines in the figure represent
respectively the four central imaging rays in four lenses of
different mass. The mass ratio of the corresponding lenses is
$2:3:4:5$ from bottom to top. We find that, when the mass $M$
increases, the central imaging ray will be closer to the lens
center.

This imaging characteristic of Fig.\ 4 can be used to explain the
missing of the central image of gravitational lensing. In practical
observations, the distances $OL$, $SL$ and the mass $M$ are all in
astronomical scale; therefore, the light of central imaging is
extremely close to the lens center. For a lens matter system denser
in the center, the possibility of the central imaging light to be
blocked increases. The relatively longer inner path of the central
imaging ray adds the possibility of scattering and absorption on the
way. Besides, the central image, if still exists, is relatively
faint compared with the bright lens core. All these factors lead to
little chance of finding the central image. This accounts for the
central image missing in almost all observed cases of gravitational
lensing.

\section{Discussion on the refraction property of the gravitational spacetime} \label{}

Though we have successfully given a refractive index description of
the gravitational lensing and solved a puzzling problem of it, there
is still a doubt about the refraction property of the gravitational
spacetime as shown in Eq.\ (4). The doubt can be eliminated if we
consider that, around the gravitational matter, there exists a
special medium --- vacuum, which may have a graded refractive index
in gravitational field.

A graded vacuum refractive index needs at least an influence of
matter or field on the vacuum. It is exhilarating to see that the
recent theoretical and experimental progresses in quantum vacuum
have provided a strong support to such an influence. Ahmadi and
Nouri-Zonoz pointed out that under the influence of electromagnetic
field, vacuum can be polarized, which has led to astonishingly
precise agreement between predicted and observed values of the
electron magnetic moment and Lamb shift, and may have effects on the
photon propagation \cite{Ahmadi2006}. Rikken and Rizzo considered
the anisotropy of the optical properties of the vacuum when a static
magnetic field $\textbf{B}_0$ and a static electric field
$\textbf{E}_0$ are simultaneously applied perpendicular to the
direction of light propagation \cite{Rikken2003}. They predicted
that magnetoelectric birefringence will occur in vacuum under such
conditions. They also demonstrated that the propagation of light in
vacuum becomes anisotropic with the anisotropy in the refractive
index being proportional to ${\textbf{B}_0}\times{\textbf{E}_0}$.
Dupays \emph{et al}. studied the propagation of light in the
neighborhood of magnetized neutron stars. They pointed out that the
light emitted by background astronomical objects will be deviated
due to the optical properties of quantum vacuum in the presence of a
magnetic field \cite{Dupays2005}.

The fact that the light propagation in vacuum can be modified by
applying electromagnetic fields to the vacuum indicates that vacuum
is actually a special kind of optical medium
\cite{Ahmadi2006,Dupays2005}. The similarity between the vacuum and
the dielectric medium implies that vacuum must also have its inner
structure, which could be influenced by matter or fields. Actually,
the structure of quantum vacuum has been investigated in quite a
number of papers recently \cite{Armoni2005,Barroso2006,Dienes2005}.

Besides the electromagnetic field, the existence of matter can also
influence the vacuum. For example, the vacuum inside a microcavity
is modified due to the existence of the cavity mirrors, which will
alter the zero-point energy inside the cavity and cause an
attractive force between the two mirrors known as Casimir effect
\cite{Gies2006,Emig2007}, which has been verified experimentally
\cite{Lamoreaux1997,Chan2001}.

The above said makes it natural to suppose that the refractive index
of vacuum can be influenced by the gravitational matter. This
thought is also supported by the theory of quantum fields, which
tells us that particles are actually the excited vacuum. And it is
unthinkable that the excited vacuum will exert no influence on the
vacuum around.

In brief, we could relate the graded refraction property of the
gravitational spacetime to the vacuum influenced by the
gravitational matter.

\section{Conclusions}
\label{}

We have shown two strong similarities between the gravitational
lensing and the conventional optical lensing: one is in the Fermat's
principle, the other is in light deflection formula. The
similarities indicate a graded refractive index analysis of the
gravitational lensing. We derived through the general relativity and
the Fermat's principle the general expression of this refractive
index for a static spherically symmetric gravitational field. From
this expression and the Schwarzschild exterior and interior
solutions, the exact refractive index profile is obtained. In weak
field case, we got a simple unified exponential function of the
gravitational potential for the calculation of the refractive index
profile outside and inside the lens matter system. Since the
derivation is based on the general relativity, there is no
inconsonance between the general relativity and our result. By using
the obtained result, we investigated the gravitational lensing in a
conventional optical way. Some results from computer simulations
interpreted the long puzzling problem of the central image missing
in a clear way. We also suggested the graded refractive index of the
gravitational spacetime be interpreted as that of the gravitational
vacuum. We hope our work will be a useful means of gravitational
lensing and be a positive stimulus to the vacuum-based investigation
of gravitation.

\ \

\appendix{\textbf{Acknowledgments}}

\

We wish to acknowledge the supports from the Ministry of Science and
Technology of China (grant no. 2006CB921403 \& 2006AA06A204) and the
Zhejiang Provincial Qian-Jiang-Ren-Cai Project of China (grant no.
2006R10025).


\appendix*{\textbf{}}
\section{Derivation of Eqs.\ (11) and (12)}

Suppose the relation between $R$ and $r$ is
\begin{equation}
\frac{dR}{f(R)}=\frac{dr}{r},
\end{equation}
where $f(R)$ is a function of $R$.

Now we consider the light path $P_\infty P_0$ in Fig.\ 5, where
$P_0$ is the closest point to the gravitational center. Combining
Eqs.\ (6), (10) and (A.1), we get
\begin{equation}
\frac{R/\sqrt{A(R)}}{f(R)}\sqrt{\left[\frac{R/\sqrt{B(R)}}{R_0/\sqrt{B(R_0)}}\right]^2-1}=\sqrt{\left(\frac{nr}{n_0
r_0}\right)^2-1}.
\end{equation}

At the infinite point $P_\infty$, the space-time becomes flat. So we
have
\begin{eqnarray}
\sqrt{A(R_\infty)}=1,\\
\sqrt{B(R_\infty)}=1,\\
\frac{dR}{dr}|_\infty=1,\ \ \\
\frac{R_\infty}{r_\infty}=1,\ \ \ \\
n_\infty=1.\ \ \ \
\end{eqnarray}

According to Eqs.\ (A.1), (A.3), (A.5) and (A.6), we have
\begin{equation}
f(R_\infty)=R_\infty/\sqrt{A(R_\infty)}.
\end{equation}

Applying Eq.\ (A.2) to the infinite point $P_\infty$ reads
\begin{equation}
\frac{R_\infty/\sqrt{A(R_\infty)}}{f(R_\infty)}\sqrt{\left[\frac{R_\infty/\sqrt{B(R_\infty)}}{R_0/\sqrt{B(R_0)}}\right]^2-1}=\sqrt{\left(\frac{n_\infty
r_\infty}{n_0 r_0}\right)^2-1}.
\end{equation}

Substituting Eqs.\ (A.4), (A.6), (A.7), (A.8) into the above
equation gives
\begin{equation}
n_0 r_0=R_0/\sqrt{B(R_0)}.
\end{equation}

Next we consider another light path $P_0 P_0^{\ '}$ in the
gravitational field. Now the closest point to the gravitational
center is not $P_0$ but $P_0^{\ '}$, then Eq.\ (A.2) becomes
\begin{equation}
\frac{R/\sqrt{A(R)}}{f(R)}\sqrt{\left[\frac{R/\sqrt{B(R)}}{R_0^{\
'}/\sqrt{B(R_0^{\ '})}}\right]^2-1}=\sqrt{\left(\frac{nr}{n_0^{\ '}
r_0^{\ '}}\right)^2-1}.
\end{equation}

Similar to Eq.\ (A.10), we can get
\begin{equation}
n_0^{\ '} r_0^{\ '}=R_0^{\ '}/\sqrt{B(R_0^{\ '})}.
\end{equation}
Applying Eq.\ (A.11) to the point $P_0$ reads
\begin{equation}
\frac{R_0/\sqrt{A(R_0)}}{f(R_0)}\sqrt{\left[\frac{R_0/\sqrt{B(R_0)}}{R_0^{\
'}/\sqrt{B(R_0^{\ '})}}\right]^2-1}=\sqrt{\left(\frac{n_0
r_0}{n_0^{\ '} r_0^{\ '}}\right)^2-1}.
\end{equation}

Combining Eqs.\ (A.10), (A.12) and (A.13) gives
\begin{equation}
f(R_0)=R_0/\sqrt{A(R_0)}.
\end{equation}

In the same way, we can obtain the following relation for every
point $P$ in the gravitational field:
\begin{equation}
f(R)=R/\sqrt{A(R)}.
\end{equation}

Substituting the above equation into Eq.\ (A.1) gives Eq.\ (11).
Combining Eqs.\ (6), (10), (A.10) and (11) gives Eq.\ (12).

\begin{figure}
\includegraphics[width=2.5in]{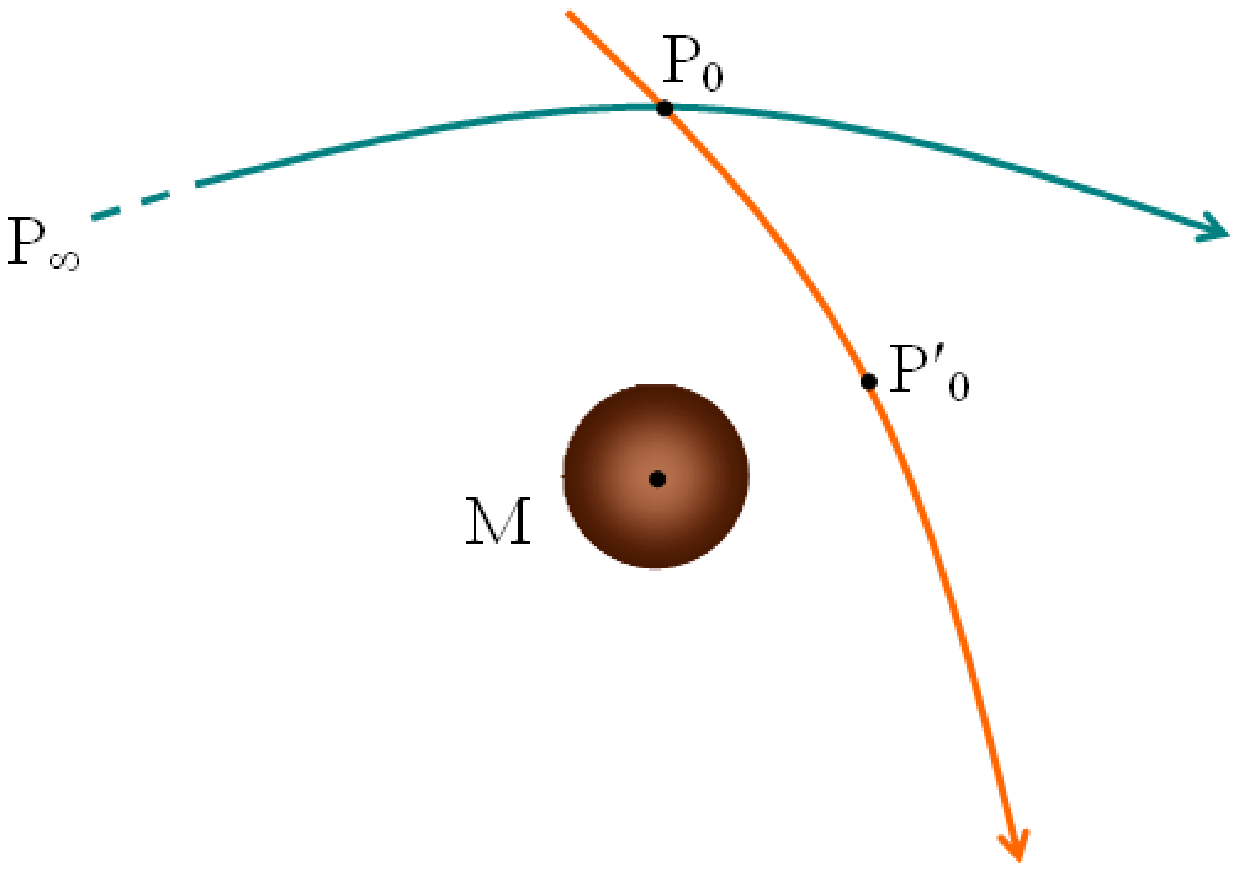}
\centering \caption{\label{fig05} Two light paths in a gravitational
field.}
\end{figure}

\newpage 

\end{document}